\documentclass[aps,
               prb,
               twocolumn,
               10pt,
               longbibliography,
               floatfix,
               nofootinbib,
               superscriptaddress]{revtex4-2}
\usepackage{cancel}
\usepackage{physics}
\usepackage[english]{babel}
\usepackage[utf8]{inputenc}
\usepackage{amssymb}
\usepackage{amsmath}
\usepackage{multirow}
\usepackage{float}
\usepackage[pdftex]{graphicx}
\usepackage{xspace}
\usepackage{dsfont}
\usepackage{soul}
\usepackage{bm}
\usepackage[normalem]{ulem}
\usepackage{hyperref}
\hypersetup{
    colorlinks,%
    citecolor=blue,%
    filecolor=blue,%
    linkcolor=blue,%
    urlcolor=blue
}
\usepackage[usenames, dvipsnames]{color}

\graphicspath{{.}{Figuras/}}

\usepackage{mathrsfs}

\begin{document}
	
\title{Emulating non-Hermitian dynamics in a finite non-dissipative quantum system}

\author{Eloi Flament$^1$, Fran\c{c}ois Impens$^2$ and David Gu\'ery-Odelin$^1$}


\affiliation{
$^1$ Laboratoire Collisions, Agr\'egats, R\'eactivit\'e, FeRMI, Universit\'e de Toulouse, CNRS, UPS, France \\
$^2$ Instituto de F\'{i}sica, Universidade Federal do Rio de Janeiro,  Rio de Janeiro, RJ 21941-972, Brazil}


	\begin{abstract}
		We discuss the emulation of non-Hermitian dynamics during a given time window by a low-dimensional quantum system coupled to a finite set of equidistant discrete states acting as an effective continuum. We first emulate the decay of an unstable state, and map the quasi-continuum parameters enabling a precise approximation of the non-Hermitian dynamics. The limitations of this model, including in particular short- and long- time deviations, are extensively discussed.  We then consider a driven two-dimensional system, and establish  criteria for the non-Hermitian dynamics emulation with a finite quasi-continuum.  We quantitatively analyze the signatures of finiteness of the effective continuum, addressing the possible emergence of non-Markovian behavior during the time interval considered. Finally, we investigate the emulation of dissipative dynamics using a finite quasi-continuum with a tailored density of states. We show on the example of a two-level system that such a continuum can reproduce non-Hermitian dynamics more efficiently than the usual equidistant quasi-continuum model.	
	\end{abstract}
	

\maketitle

\section{Introduction}

The decay of unstable states occurs in a wide range of  areas of quantum mechanics,  including 
atomic physics with the limited lifetime of excited electronic states in atoms, condensed matter with various relaxation processes in quantum dots electronic states, in polaron or exciton physics to name but a few, nuclear physics with the exponential decay law in radioactivity, or high energy physics with the short lifetime of particles such as the Higgs boson.   The basic phenomenon underlying these decays is fundamentally the same. It is the irreversible transition from an initial unstable state to a continuum of final states. Such a decay can be derived from first principles.  In the perturbative limit, this problem often offers a first introduction to open quantum systems with the Fermi's golden rule. 
Besides the perturbative limit, the complete resolution of the model reveals three different successive regimes characterized by different decay laws \cite{1958JETP1053K,PhysRevD.16.520,Greenland}: on very short time \cite{raizenshorttime}, the decay is quadratic, it is subsequently governed by an exponential law at intermediate time, and eventually exhibits a power law tail on long time scales \cite{PhysRevLett.96.163601,PhysRevA.80.012703}.  In general, these studies reveal that decay can be sensitive to the structure of the environment.

Quantum simulations have become a very important research topic, with various fundamental and technological applications~\cite{QuantumSimulatorRoadmap}. As any realistic quantum process involves a finite amount of dissipation, a quantum decay emulator appears as an interesting building block for such systems. The simplest model of quantum decay corresponds to the inclusion of a non-Hermitian contribution to the Hamiltonian, which suggests emulating non-Hermitian systems. Non-Hermitian dynamics also have their own interest. Since the realization of complex optical PT potentials~\cite{NonHermitianTransport0,ComplexPotential2}, the community has unveiled a very rich phenomenology and numerous applications for effective non-Hermitian systems. To name a few, we can mention the non-Hermitian skin effect~\cite{SkinEffect1,SkinEffect2}, the non-Hermitian transport~\cite{NonHermitianTransport0,NonHermitianTransport4,Damanet19,NonHermitianTransport4bis}, and more generally the intriguing topology of effective non-Hermitian systems~\cite{NonHermitian1,NonHermitian2,NonHermitian3,NonHermitian4,NonHermitian5,NonHermitian6,NonHermitian7}. Emulating non-Hermitian dynamics can provide access to the above phenomena in different platforms.
	
Engineering a truly non-Hermitian and irreversible quantum dynamics over an arbitrarily long time requires an interaction of the system with an infinite set of states, as in the usual paradigm of infinite discrete quasi-continuum~\cite{STEY19721,LivreClaude}. Nevertheless, the emulation of quantum dissipation during a finite time can be sufficient for experimental purposes  - for instance when dissipation is used as an asset to prepare a given quantum state~\cite{DissipativeQuantumStatePreparation}. In this context, simulating dissipative quantum dynamics thanks to a coupling with a finite - and ideally minimal - number of ancilla states seems a feasible task. This possibility may have interesting applications in quantum computing, where a smaller number of ancilla states usually corresponds to a simpler setup.

\begin{figure}[t!]
 	\centerline{\includegraphics[scale=0.7]{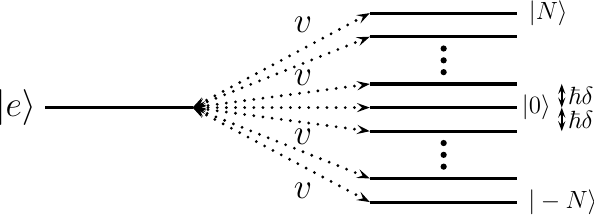}}
 	\caption{Schematic picture of a quantum system (consisting here in a single discrete state $| e \rangle$) coupled to a finite set of equidistant discrete levels. This model mimics the coupling to a continuum. }
 	\label{figureI1}
 \end{figure}

The purpose of this article is to investigate this avenue and provide an emulation of non-Hermitian dynamics for a given time interval with a quasi-continuum made of a \textit{finite} set of ancilla states (see Fig.~\ref{figureI1}). We use the trace distance to quantify the quality of our model, and discuss in detail the minimum number of levels required to obtain an accurate emulation. We also investigate separately the short- and long-term behavior of the associated dynamics.  At early times, we compare the quantum evolution of the coupled system with the Zeno effect expected from a genuine continuum. At long times, we observe and characterize quantitatively the emergence of revivals in the presence of the finite continuum, enabling us to set an upper limit on the validity time of this emulation. We connect the appearance of these revivals with adequate measures of non-Markovianity.

We proceed as follows. In Sec.~\ref{sec1}, we provide a brief reminder on the decay for a single discrete level coupled to an infinite continuum. Section \ref{Sec2} presents the considered quasi-continuum model, composed of equidistant energy levels equally coupled to a given state, and discuss its main features.  In Sec.~\ref{Sec3}, we investigate the same issues for a two-level system whose excited state is coupled to a continuum. We identify a method for defining the minimum size of the discrete continuum using a Fourier analysis. In Sec.~\ref{Sec4}, we discuss the emergence of a non-Makovian evolution at long times, and build up on the previous sections to design a discrete quasi-continuum with a minimum number of states to reproduce the expected behavior in the strong coupling limit.

\section{Decay of a single level coupled to an effective continuum}
\label{sec1}

 We illustrate our method by first considering a system consisting of a single eigenstate $|e\rangle$ coupled to a large set of independent states $\{ | \varphi_f \rangle \}$.  This system is the usual paradigm explaining the irreversible exponential decay and  Lamb shift undergone by a quantum state coupled to a continuum~\cite{LivreClaude}. We briefly recall below the corresponding derivation in the standard case of an infinite and broad effective continuum consisting of the set of states $\{| \varphi_f \rangle\}$. The quantum system under consideration follows an Hamiltonian given by the sum $H=H_0+V$ with $H_0= E_e |e  \rangle \langle e | + \sum_{f} E_f | \varphi_f \rangle \langle \varphi_f | $ the free-system Hamiltonian diagonal in the basis $\{|e \rangle, | \varphi_f \rangle \}$, and with the off-diagonal contribution $V= \sum_f V_{fe} |\varphi_f  \rangle \langle e |+h.c.$ accounting for the coupling between the discrete state and the effective environment.

We search for a solution of the time-dependent Schr\"odinger of the form:
\begin{equation}
|\psi(t)\rangle=c_e(t)e^{-iE_et/\hbar}|e\rangle+ \sum_f
c_f(t)e^{-iE_ft/\hbar}|\varphi_f\rangle.    
\end{equation}
and subsequently obtain by projection on the eigenstates of $H_0$ the following integro-differential obeyed by the coefficient $c_e$:
\begin{equation}
	\label{eqce}
   \dot{c}_e(t)=-\int_0^tdt'K(t-t')c_e(t'), 
\end{equation}
where the kernel is defined by
\begin{equation}
  K(\tau)= \frac{1}{\hbar^2} \sum_f |V_{ef}|^2e^{i \omega_{ef}\tau},
  \label{eqk}
\end{equation}
with $\omega_{ef}=(E_e-E_f)/\hbar$. Equations (\ref{eqce},\ref{eqk}) capture the exact quantum dynamics of this system and so far involve no assumptions about the set of final states $ \{ | \varphi_f \rangle \}$. The function $K(\tau)$ accounts for the memory of the effective environment, resulting in a possibly non-Markovian evolution for the amplitude $c_e(t)$.

 We now assume that the effective continuum $\{ | \varphi_f \rangle \}$ covers a wide range of frequencies. As a result, the $K(\tau)$ function is expected to be sharply peaked around $\tau=0$ when compared to the time-scale of the amplitude evolution - for a genuine continuum with a flat coupling, the sum over all possible final states in Eq.~\eqref{eqk} would actually yield a Dirac-like distribution. This large timescale separation enables one to pull out  the amplitude $c_e(t)$ from the integration of the memory kernel  in Eq.~\eqref{eqce}, and to extend the boundary of this integral to infinity. We then obtain a simple closed differential equation for $c_e$: 
\begin{equation}
	\label{eqce2}
   \dot{c}_e(t)=-\left( \int_0^\infty d\tau K(\tau) \right)c_e(t). 
\end{equation}
The pre-factor is readily derived in the framework of complex analysis:
\begin{equation}
	\label{eq:KernelIntegral}
    \int_0^\infty d\tau K(\tau) = i\Delta \omega_e +  \frac{\Gamma}{2}, 
\end{equation}
with
\begin{eqnarray}
    \frac{\Gamma}{2}   & = &   \frac{2 \pi}{\hbar} \sum_f |V_{ef}|^2 \delta (E_e-E_f)  \nonumber \\
    \Delta \omega_e & = & \frac{1}{\hbar} {\cal P} \left(\sum_f \frac{|V_{ef}|^2}{E_e-E_f} \right),  \label{eq:Lambshift}  \label{eq6}
\end{eqnarray}
where ${\cal P}$ denotes the principal value. For the considered coupling to a large set of states, the main effects on the discrete state are therefore an exponential decay of the population at a rate $\Gamma$ witnessing an irreversible evolution as well as a frequency shift $\Delta \omega_e$ commonly referred to as the Lamb shift. Equation~\eqref{eq6} simply expresses Fermi's golden rule for the effective continuum with the density of states $\rho(E)= \sum_f \delta (E-E_f) $. Remarkably, Fermi's golden rule holds not only for a genuine continuum but also for a countable set $\{| \varphi_f \rangle\}$ involving only discrete states~\cite{LivreClaude}. Finally, unlike Eq.~\eqref{eqce}, the amplitude $c_e(t)$ at a given time no longer depends on its history: the effective continuum $\{ | \varphi_f \rangle \}$ behaves as a Markovian environment. Eqs.(\ref{eqce2},\ref{eq:KernelIntegral}) implicitely define an effective non-Hermitian Hamiltonian $H_{\rm eff}=  \Delta \omega_{ge} - i  \frac{\Gamma}{2}$ for this one-level system.

A closer look at Eqs.~(\ref{eq6}) reveals the central role played by the density of states $\rho(E)$ of the effective continuum \cite{PhysRev.123.1503,Fonda_1978,PERES198033,Greenland}. Indeed, its properties are responsible for deviations to the exponential law both at short and long times: the existence of an energy threshold ($\rho(E<E_0)=0$) generates long-time deviations while finiteness of the mean energy ($\int \rho(E) EdE< \infty$) explains the short time deviations. 

In the same spirit, we examine below how the two characteristics of the quantum evolution discussed above - exponential decay and non-Markovianity - are affected by the use of a finite set as an effective continuum. We restrict our attention to a finite time-interval, as only infinite sets may reproduce these characteristics during an arbitrary long times.

\section{Coupling of a single state to a finite discretized continuum}
\label{Sec2}

\textit{Description of the FQC model}. To quantitatively characterize such an irreversible process, we introduce a Finite Quasi-Continuum (FQC) model consisting of a finite set of equidistant energy levels, which are equally coupled to a given state $|e\rangle$ (See Fig.~\ref{figureI1}). This system mimics the decay of an unstable discrete state $|e\rangle$ in a finite time window. In what follows, unless otherwise stated, we always consider FQCs composed of $N_{\rm FQC}=2N+1$ equidistant energy levels symmetrically distributed around the unstable  state energy -- set by convention to $E=0$. Here, the total Hilbert space is of dimension $N_{\rm tot}= N_{\rm sys}+ N_{\rm FQC}=2N+2$. We denote $\hbar \delta$ the energy gap between two successive FQC states and $v=|V_{fe}|$ the flat coupling strength between the FQC and the discrete state $|e\rangle$. The expected decay $\Gamma$ in the limit $N \rightarrow + \infty$ is given by Eq.~\eqref{eq6} which captures the dynamics of an infinite discrete continuum, namely
\begin{equation}
	\Gamma=\frac{2\pi}{\hbar^2}  \frac{v^{2}} {\delta} \label{eq1}.
\end{equation}
which corresponds to the Fermi's golen rule. In the following, we consider FQCs associated with a fixed common decay rate $\Gamma$. We therefore impose $v^2 / \delta = Cte$. In our numerical resolution, we implicitly normalize the energies by $\hbar \Gamma$ and the time by $\Gamma^{-1}$ which amounts to taking $\hbar=1$ and $\Gamma=1$. Our results are valid for arbitrary values of the dissipation rate $\Gamma$ as long as the dimensionless parameters $\overline{v}=v/( \hbar \Gamma)$, $\overline{t}_f= \Gamma t_f,$... remain identical. The considered FQCs are therefore entirely determined by their size ($2N+1$) and the coupling strength, $v$.

  The model Hamiltonian reads in matrix form
\begin{equation}
H=
\begin{pmatrix}
 0  & v & ... & v & v & v & v\\
v  & -N \hbar \delta & 0 &  ... & 0 & 0 & 0\\
v  & 0 & -(N-1)\hbar \delta & 0 & ... & 0 & 0\\
v & 0 & ... & 0 & ... & 0  & 0\\
v & 0 & . & . & 0 & (N-1) \hbar \delta & 0\\
v & 0 & 0 & 0 &... & 0 & N \hbar \delta\\
\end{pmatrix}.
\label{Ha1}
\end{equation}

\begin{figure}[t!]
    \centerline{\includegraphics[scale=0.5]{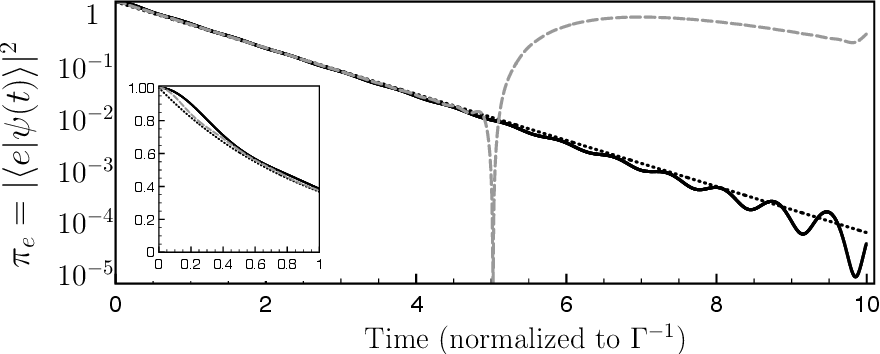}}
    \caption{FQC vs genuine continuum for a single discrete level: Excited state population $\pi_e(t)$ of a discrete level coupled to FQCs obtained from the full quantum dynamics under the Hamiltonian~\eqref{Ha1} with $N=15$ and $v=0.3 \: \hbar \Gamma$ (solid black line) and $N=15$ and $v=0.45 \: \hbar \Gamma$ (gray dashed line) as a function of time (normalized to $\Gamma^{-1}$). The dotted line represents the exponential decay expected from a genuine continuum.}
    \label{figureI2}
\end{figure}

\textit{Examples of FQCs and connection with the Zeno effect}. In Fig.~\ref{figureI2}, we compare the evolution of the excited state population for an example of FQC (solid black line) with the exponential decay expected from  Fermi's golden rule (dotted line). As expected, we observe a very good agreement, with minor discrepancies at short times (see the inset of Fig.~\ref{figureI2}) and at long times when the population is extremely small. We used a FQC with $N=15$ and a coupling strength $v=0.3 \: \hbar \Gamma$. In this case, emulation of quantum decay does not require a very large Hilbert space.

The disagreement at short times corresponds to a quadratic decay of the excited state coupled to a FQC. The initial quadratic profile is directly related to the Zeno effect. It is found by expanding the evolution operator for a short amount of time $\delta t$, by writing $|\psi(\delta t) \rangle =e^{-iH\delta t/\hbar} |e \rangle  \simeq |e \rangle + |\delta \psi \rangle$ with 
\begin{equation}
   |\delta \psi \rangle =  \left( - \frac {i H} {\hbar}  \delta t - \frac{H^2}{2 \hbar^2} (\delta t)^2 \right)  |e \rangle.
\end{equation}
We infer the initial state population $\pi_e( t)=|\langle e |\psi( t) \rangle |^2$ at early times
\begin{equation}
    \pi_e( \delta t) \simeq 1 - \frac{(\delta t)^2}{T_Z^2},
\end{equation}
where $T_Z^{-2} = \frac {1} {\hbar^2}  (\langle H^2\rangle_e-\langle H \rangle_e^2)= \frac {1} {\hbar^2}   \sum_n \langle e | V | n \rangle \langle n | V | e \rangle = (2N+1)   ( v / \hbar)^2$. The duration $T_Z$ corresponds to the Zeno time and decreases with the size of the FQC. As $T_Z$ vanishes in the limit $N \rightarrow + \infty,$ the observed initial quadratic profile witnesses the limited number of states of the FQC. For the parameters $N=15$ and $v=0.3 \: \hbar \Gamma$, one finds $T_Z \simeq 0.6 \: \Gamma^{-1}$, consistent with the inset of Fig.~\ref{figureI2}

We now provide a second example of FQC for which the excited state population evolves very differently from the expected exponential decay. We take a FQC with $N=15$ and $v=0.45 \: \hbar \Gamma$, which corresponds to a larger energy gap between the FQC levels than in the first example, and therefore further away from an ideal continuum. Good agreement is observed up to $t\simeq 5 \Gamma^{-1}$, when the population $\pi_e(t)$ grows abruptly (gray dashed line, Fig.\ref{figureI2}). This revival of the probability distribution in the discrete state unveils the underlying fully coherent dynamics.

\begin{figure}[t!]
    \centerline{\includegraphics[scale=0.56]{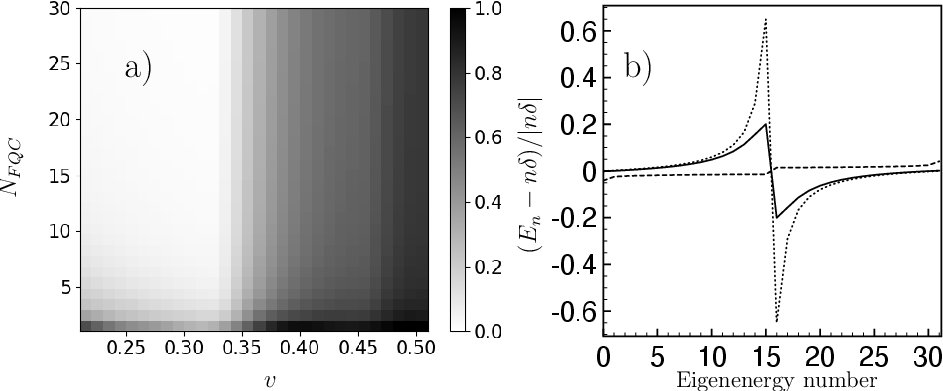}}
    \caption{(a) Quality of the emulation of a continuum by a FQC: Parameter $\mathcal{D}_1$ (obtained by a numerical resolution of the Shcr\"odinger equation) as a function of the FQC parameters $\{ N, v \}$ ($v$ is given in units of $ \hbar \Gamma$). The white zone reveals a very good agreement with the exponential decay expected from a genuine continuum. b) Spectrum analysis of the Hamiltonian~\eqref{Ha1}: Eigenenergies $E_n$ in crescent order normalized by $\delta$ for $N=15$ and $\delta = 5 \: \Gamma$ (dotted line), $\delta = 0.5 \: \Gamma $ (solid line) and $\delta = 0.05 \: \Gamma$ (dashed line). $N_{FQC}=2N+1$ is the FQC size.}
    \label{figureI3}
\end{figure}

\textit{Quantitative mapping of successful FQCs for the emulation of single-state decay}. We now proceed to a quantitative mapping of the FQC parameters $(N,v)$ suitable for accurate continuum emulation.  In order to capture the accuracy of our model on a given time window, one needs a distance measure between the quantum evolution observed in the presence of an FQC and the genuine continuum respectively. For the single-state quantum system considered here, the density matrix boils down to the excited state population $\pi_e(t)$. We therefore introduce the following distance
\begin{equation}
	\label{eq:D1}
\mathcal{D}_1(t_f) = \frac{1}{t_f} \int_0^{t_f} |\pi_e(t) - \pi_0(t)| dt.
\end{equation}
as a figure of merit for the quality of the FQC emulation over the time window $0 \leq t \leq t_f$. $\pi_0(t)=\pi_e(0)e^{-\Gamma t}$ is the exponential decay expected in the large continuum limit. We choose $t_f$ to be larger than several $\Gamma^{-1}$ to account at best for the full decay. In our numerical examples, we systematically use $t_f=10 \Gamma^{-1}$ (unless otherwise specified). The results are summarized in Fig.~\ref{figureI3}a. The good set of parameters for the chosen time interval is provided by the white area. This figure reveals that the quality of the emulation increases with the number of FQC states and decreases with the potential strength $v$ - corresponding to FQCs with a larger energy gap $\hbar \delta$ for a fixed decay rate $\Gamma$. In particular, the quality of the emulation  drops off sharply above a critical coupling value $v_c\simeq 0.32 \: \hbar \Gamma$, which is independent of the number of FQC states. We explain below this abrupt change in terms of quantum interference and revivals of the discrete state population. The dashed gray line of figure~\ref{figureI2} provides an example of revival of the excited population $\pi_e(t)$ coupled to a FQC with a strength $v \geq v_c$.

\begin{figure}[h!]
    \centerline{\includegraphics[scale=0.57]{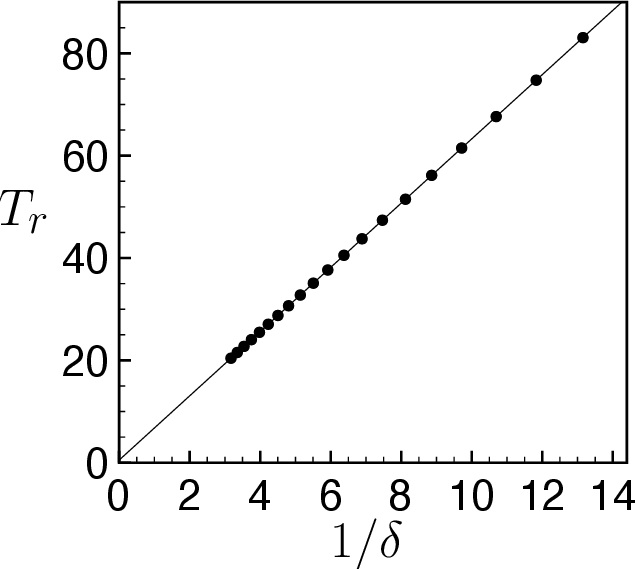}}
    \caption{Black disks: Resurgence time $T_r$  as a function of the inverse of the FQC energy gap $\delta$. $T_r$ is obtained by a numerical resolution of the Schr\"odinger equation with the Hamiltonian (\ref{Ha1}). The solid black line represents a linear fit $T_r = a /\delta$ yielding $|a-2\pi| \leq 10^{-3},$ showing thus an excellent agreement with our prediction for the revival time.}
    \label{figres}
\end{figure}

We now provide a quantitative analysis of the occurence of such revivals in a given time window. We first look for a necessary condition of revival. For this purpose, we expand the wavefunction at time $t$ on the eigenbasis:
\begin{equation}
	|\psi(t) \rangle = \sum_{n=0}^{N_{\rm tot}} a_n(0)e^{-iE_nt/ \hbar} | \psi_n\rangle,
\end{equation}
where $E_n$ are the eigenenergies of the total Hamiltonian~(\ref{Ha1}) and with $N_{\rm tot}=2N+2$ the dimension of the total Hilbert space. We denote by $T_r$ the revival time, which necessarily fulfills $||   |\psi(T_r)\rangle - |\psi(0)\rangle ||^2  \ll 1,$ namely 
\begin{equation}
 \sum_{n=0}^{N_{\rm tot}} |a_n(0)|^2 (1- \cos(E_n T_r/\hbar))   \ll \frac 1 2.
	\label{resurg} 
	\end{equation}
The revivals correspond to a constructive quantum interference occurring at a time $T_r$ determined by the Hamiltonian~(\ref{Ha1}) spectrum. Actually, this spectrum is only marginally affected by the coupling to the discrete state, and has a nearly linear dependence of its eigenvalues $E_n \simeq n \hbar \delta$ (see the numerical analysis on Fig.~\ref{figureI3}b). This results is valid for a wide range of energy gaps $\delta$. The condition~\eqref{resurg} requires that for all values of $n$,  $E_n T_r/\hbar =2\pi k_n$ with $k_n$ an integer. As $E_n \simeq n \hbar \delta$, we find $k_n=n$ and $T_r = 2 \pi  / \delta$. Figure~\ref{figres} confirms numerically the predictions of this simple revival model. We have plotted the revival time inferred from the exact resolution of the Schr\"odinger equations of the model with the Hamiltonian (\ref{Ha1}) as a function of $1/\delta$.

The above analysis provides a clear criterion for the suitability of the FQC for emulating irreversible dynamics. A necessary condition is the absence of revival during the considered time windows, i.e. $t_f < T_r$. This sets an upper bound on the energy gap, namely $\delta \leq \delta_c=2 \pi/t_f$, or equivalently on the coupling strength $v \leq v_c  = \hbar \sqrt{\Gamma/t_f}$ as both quantities are related by Eq.~\eqref{eq1}. For the considered final time $t_f=10 \: \Gamma$, we obtain the value $v_c=0.316 \: \hbar \Gamma$ in very good agreement with the numerical results of Fig.~\ref{figureI3}a. The region $v \geq v_c$ corresponds indeed to the onset of the gray zone accounting for a degradation in the emulation of dissipative dynamics. In the next Section, we investigate the appropriate choice of the FQC model parameters in the different regimes of a driven two-level system.

\section{Coupling of a two-level system to a finite discretized continuum}
\label{Sec3}

\begin{figure}[t!]
	\centerline{\includegraphics[scale=0.70]{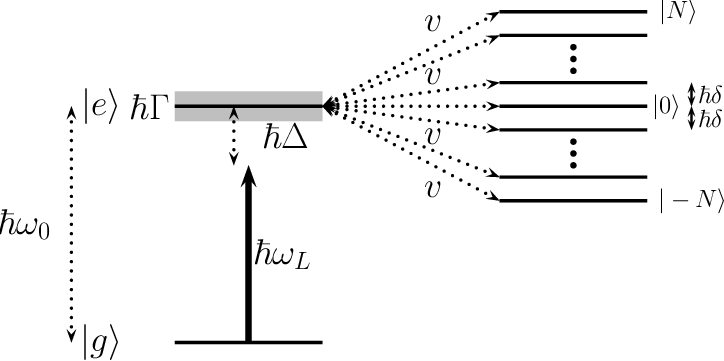}}
	\caption{Two-level system driven by a laser pulse with a detuning ($\Delta$), involving a stable ground state $|g \rangle$ and an excited state $|e \rangle$ coupled to a large but finite set of discrete levels. This coupling emulates an unstablity and yields an effective linewidth $\Gamma$ for the transition. }
	\label{figureII1}
\end{figure}

\textit{Model description and equations of motion}. In this Section, we consider a two-level atom with a stable ground state $|g\rangle$ and an unstable excited state $|e \rangle$ (see Fig.~\ref{figureII1}), which is the standard model for spontaneous emission in quantum optics~\cite{LivreClaude}. We denote $\omega_0$ the transition frequency of this two-level system, and assume that it is illuminated by a nearly-resonant laser of frequency $\omega_L \simeq \omega_0$. This external field drives the system with a Rabi coupling of frequency $\Omega_0$ between the two atomic levels. The excited state acquires a finite width $\Gamma$ due to its coupling with the continuum. 

We now consider a $N_{\rm tot}=2N+3$-dimensional Hilbert space encapsulating the two-level quantum system and the FQC. Considering the driving term, the total Hamiltonian is given by 
\begin{equation}
	H= 
	\begin{pmatrix}
		0 & \hbar\Omega_0 & 0 & 0 & ... & 0 & 0\\
	\hbar	\Omega_0 & \hbar \Delta & v & v & ... & v & v\\
		0 & v & -N \hbar\delta & 0 & 0 & ... & 0\\
		0 & v & 0 & -(N-1) \hbar \delta & 0 & ... & 0\\
		0 & . & . & 0 & . & 0 & ...\\
		0 & . & . & . & 0 & . & 0\\
		0 & v & 0 & 0 &... & 0 & N \hbar\delta\\
	\end{pmatrix}.
	\label{Hal2}
\end{equation}
 in the basis $\{ | g \rangle, | e \rangle, | \psi_f \rangle \}$ transformed in the rotating frame with the detuning $\Delta = \omega_0 - \omega_L$. For a given dissipation rate $\Gamma$, the system is therefore determined by four independent
 driving ($\{\Omega_0, \Delta \}$) and  FQC ($\{N,\delta \}$, or equivalently $\{N,v \}$ from Eq.\eqref{eq1}) parameters. We denote $| \psi \rangle$ the quantum state of the full Hilbert space. The corresponding density matrix $\rho= | \psi \rangle \langle \psi |$ follows a unitary dynamics $i \hbar \frac {d \rho} {dt} = [H,\rho].$ We now focus on the non-unitary quantum dynamics in the reduced Hilbert space. Specifically, we consider the evolution of the 
 $2\times2$ density matrix $\rho_r= P_{ge} \rho  P_{ge}$, where  $ P_{ge}=|g \rangle\langle g | + |e \rangle\langle e |$ is the projector on the two-dimensional Hilbert space of the system. The reduced density matrix $\rho_r$ can be obtained by first solving the full unitary dynamics and then applying the projector. In order to highlight the role played by the FQC, the equation of motion for the reduced density matrix can be rewritten in the following form
\begin{equation}
	\label{eq:reducedFQC}
	i \hbar \frac{d  \rho_r}{dt}=[H_0,\rho_r]+ S_r^{\rm FQC}.
\end{equation}
The r.h-s contains the unitary driving by the system Hamiltonian $H_0= P_{ge} H P_{ge}$ as well as a source term accounting for the interaction with the FQC  
\begin{equation}
	\label{eq:sourceFQC}
S_r^{\rm FQC}=\left(
	\begin{array}{ll}
		0 & \lambda_N \\
		\lambda_N^* & \eta_N
	\end{array}
	\right).
\end{equation}
where $ \lambda_N=v\sum_{i=2}^{2N+2} \rho_{gi}$ and $\eta_N= v\sum_{i=2}^{2N+2} (\rho_{ie}-\rho_{ei})$. This source term drives effective non-unitary dynamics in the considered time interval, and depends on the coherences between the FQC levels and the quantum system. The equations above contain no approximation and capture the full quantum dynamics of the 2-level system coupled to a FQC.

\begin{figure}[h!]
 	\centerline{\includegraphics[scale=0.70]{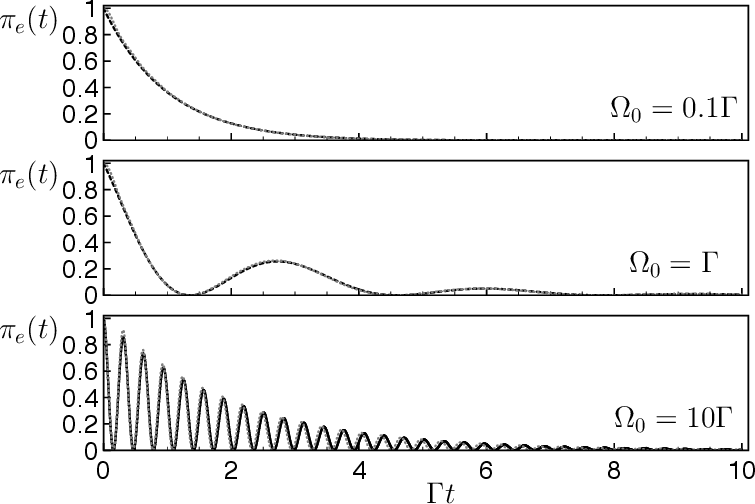}}
 	\caption{Non-Hermitian vs FQC dynamics for the two-level system in the following regimes: over-damping ($\Omega_0 =0.1 \Gamma$, upper panel), critical  ($\Omega_0 =1 \Gamma$,middle panel) and under-damping ($\Omega_0 =0.1 \Gamma$, lower panel) in the presence of a FQC with $N_{\rm FQC}=2N+1=61$ levels and $v=0.3 \: \hbar \Gamma$ (full quantum dynamics, gray dotted line) or from the non-Hermitian dynamics (Eq.\eqref{eqcenh}, black dashed line). Both lines are superimposed, showing an excellent emulation of non-Hermitian dynamics with the considered FQCs.}
 	\label{figure6}
 \end{figure}

\textit{Non-Hermitian dynamics}. Here, we briefly review the equations of motion under an effective non-Hermitian Hamiltonian. Beyond their applications in nanophotonics, effective non-Hermitian Hamiltonians adequately describe the dynamics of open quantum system in many experimental situations. For instance, this approach has been successfully used to explain subradiance effects in large atomic clouds~\cite{Bienaime12}. Here, similarly as in Section~\ref{sec1}, the effective non-Hermitian Hamiltonian is obtained by deriving the differential equations for the two-level system probability amplitudes $(c_e,c_g)$. Using the rotating wave-approximation, one finds $H_{\rm eff}=H_0+iH_d$ with $H_0= \hbar \Omega_0 (|e \rangle \langle g| + h.c.) + \hbar \Delta | e \rangle \langle e |$ and $H_d= - \frac  {\hbar} {2} \Gamma | e \rangle \langle e |$. The anti-Hermitian contribution $iH_d$ captures the decay towards the continuum.  The evolution of the reduced density matrix under the influence of this effective Hamiltonian takes a form analogous to Eq.~\eqref{eq:reducedFQC}
\begin{equation}
	\label{eq:nonHermitianexact}
	i\hbar \frac{d\tilde \rho_r}{dt}=[H_0,\tilde \rho_r]+ S_r^{\infty}
\end{equation}
with a source term $S_r^{\infty}=i [H_d, \tilde \rho_r]_+$ capturing the non-unitary dynamics ($[ \: ]_+$ is an anti-commutator). Numerical analysis confirms that $S_r^{\infty}$ also corresponds to the limit of the FQC source terms $S_{r}^{\rm FQC}$~\eqref{eq:sourceFQC} in the large quasi-continuum limit $N \rightarrow + \infty$. At resonance ($\Delta =0$), the Schr\"odinger equation in the presence of $H_{\rm eff}$ boils down to the equation of a damped harmonic oscillator
 for the probability amplitude $c_e$
 \begin{equation}
 	\ddot c_e + \Gamma \dot c_e/2 + \Omega_0^2 c_e=0.
 	\label{eqcenh}
 \end{equation}
One identifies the three usual dynamical over/critical/under-damping regimes determined by the ratio $\Omega_0/\Gamma$  (see the black dashed lines in Fig.~\ref{figure6}).

\textit{Example of successful FQC-emulated dynamics}. In Fig.~\ref{figure6}, we investigate the suitability of a FQC with parameters $\{ N,v \}=\{ 30,0.3 \: \hbar \Gamma \}$ for the emulation of non-Hermitian dynamics in these different regimes. We obtain the evolution of the excited state population $\pi_e(t)$ coupled to this FQC by a numerical resolution of the Schr\"odinger equation with the Hamiltonian~\eqref{Hal2}, and compare it to the evolution under the non-Hermitian dynamics given by Eq.~\eqref{eqcenh}. Excellent agreement is observed in the three distinct regimes, covering a wide range of $\Omega_0/\Gamma$ values. We investigate below how to determine the minimal number of levels of an adequate FQC.

\textit{Quantitative mapping of successful FQCs for the emulation of two-level non-Hermitian dynamics}. Before proceeding to a more systematic analysis of the FQC suitability, we introduce a quantitative measure for the accuracy of FQC-emulated dynamics. Specifically, in the considered two-level system, we take the trace distance~\cite{NielsenChuangBook} between the reduced density matrices evolved respectively under the influence of a FQC (unitary evolution with $H$~\eqref{Hal2} followed by projection with $P_{eg}$) and following  non-Hermitian dynamics (Eq.(\ref{eq:nonHermitianexact})). This distance is defined for two density matrices $\rho$ and $\sigma$ by
	\begin{equation}
		T(\rho, \sigma)=\frac{1}{2} \mbox{Tr}\bigg(\sqrt{(\rho-\sigma)^{\dagger}(\rho-\sigma)}\bigg).
		\label{eq4}
	\end{equation}
In order to obtain a quantitative estimate of the fidelity over the whole considered interval, we use the mean trace distance over the considered time window:
	\begin{equation}
		\label{eq:meantracedistance}
\mathcal{D}_2 (t_f) = \frac{1}{t_f} \int_0^{t_f} T(\rho_e(t), \sigma(t)) dt.
\end{equation}
This definition in terms of trace distance coincides with the measure $\mathcal{D}_1$ introduced in Eq.~\eqref{eq:D1} in the one-dimensional case.

\begin{figure}[h!]
    \centerline{\includegraphics[scale=1.]{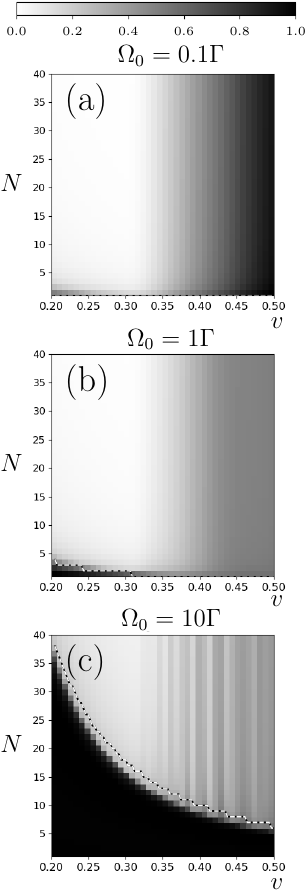}}
    \caption{Quality of the emulation of non-Hermitian dynamics with a FQC: 2D plots of the mean trace distance (normalized to its maximum value) between the non-Hermitian model and the dynamics in a FQC model with parameters $N_{\rm FQC}=2N+1$, $v$ (in units of $\hbar \Gamma$) for the respective ratios $\Omega_0/\Gamma=0.1 ({\rm a}) ,1 ({\rm b}),10 ({\rm c})$. The dotted black-white line corresponds to the number $N_{\rm max}(v)$.}
    \label{figureII3}
\end{figure}

As in Section~\ref{Sec2}, we proceed to a systematic study of the appropriate FQC parameters $(N,v)$ for the emulation of non-Hermitian dynamics. We consider here separately the three different regimes evidenced by Eq.~\eqref{eqcenh} and we use the mean trace distance~\eqref{eq:meantracedistance} between the respective density matrices evolving in the presence of a FQC ($\rho_r$) or following non-Hermitian dynamics ($\tilde \rho_r$). The results are summarized in Figs.~\ref{figureII3}(a,b,c) for the different ratios $\Omega_0/\Gamma$ corresponding to the three distinct regimes of non-Hermitian dynamics. In order to avoid the revival effect discussed in Section~\ref{Sec2}, we take a slightly shorter time interval $t_f=8 \Gamma^{-1}$. A comparison between the mappings presented Fig.~\ref{figureI3}a and Fig.~\ref{figureII3}(a,b,c) reveals very different characteristics in the FQC emulation for the one- and two-level systems. For the one-level system, successful FQC emulation only requires the absence of revivals - associated with a condition $v \leq v_c$ independent of the  FQC size $N$. Differently, we see for the two-level case that the number $2N+1$ of FQC states has a critical influence on the fidelity of the FQC-emulated dynamics. These figures reveal an abrupt transition when the parameter $N$ falls below a critical value $N(v)$ depending on the coupling strength $v$ for a given ratio $\Omega_0/\Gamma$. This raises the question of how to choose the suitable FQC parameters.

\textit{Suitability criteria for FQC}. Here we determine the subset of FQC states that are significantly populated during time evolution. Intuitively, this set should form the minimal FQC which accurately captures dissipative quantum dynamics. As can be seen below, the populated modes depend essentially on the Rabi frequency $\Omega_0$ and dissipation rate $\Gamma$.  

 This situation is reminiscent of the dynamical Casimir effect~(DCE), in which a continuum of vacuum electromagnetic modes becomes gradually populated under the harmonic motion of a moving mirror (See Ref.~\cite{DynamicalCasimirReview} for a review). In the DCE, the mirror oscillation at a frequency $\Omega_m$ induces the emission of photons of frequencies $\omega \leq \Omega_m$ in initially unpopulated electromagnetic modes. A similar effect is observed with a moving two-level atom~\cite{Souza18,Impens22} in the vacuum field.  We find below that our FQC model with a Rabi driving reproduces these features, with the emergence of sidebands at the Rabi frequency in the FQC population. As in the DCE, the external drive provides energy to the system, which eventually leaks to the continuum.

To analyse this effect, we introduce the expansion
\begin{equation}
 | \psi(t) \rangle = c_e(t) |e \rangle + c_g(t) | g \rangle + \sum_{p=-N}^N c_p(t) | p \rangle  
 \label{eq22}
\end{equation}

into the Schr\"odinger equation. A projection on the $k^{th}$ state of the FQC yields a differential equation for the coefficient $c_k(t)$ driven by the excited state probability amplitude $c_e(t).$ This equation is formally solved as
\begin{equation}
    c_k(t) = - \frac {i v} {\hbar} \int_0^t c_e(t') e^{ik\delta t'} dt'.
\end{equation}

\begin{figure}[t!]
    \centerline{\includegraphics[scale=0.70]{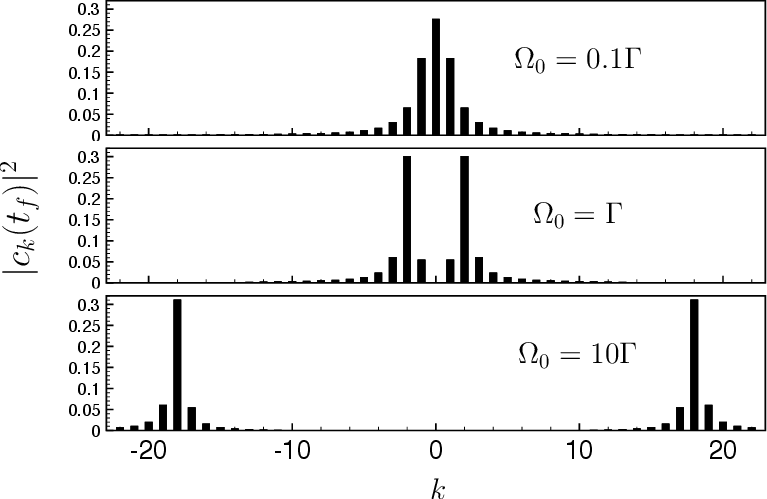}}
    \caption{Distribution of populations:  Occupation probability $|c_k(t_f)|^2$ at the final time $t_f$ as a function of the level $k$ of the FQC for $\Omega_0 =0.1 \: \Gamma$ (lower panel),  $\Omega_0 = \Gamma$ (middle panel) and $\Omega_0 =10 \: \Gamma$ (upper panel). Parameters: $2N+1=45$ levels, and with $v=0.3 \: \hbar \Gamma$.}
    \label{figure8}
\end{figure}

In the long-time limit, the coefficient $c_k(t)$ tends towards the Laplace transform of the excited state amplitude at the frequency $k \delta$ (up to a constant factor). In order to estimate the occupation probability $|c_k(t)|^2$ at time $t<t_f$, we use the probability amplitude $\tilde{c}_e(t)$ given by non-Hermitian dynamics (Eq.\eqref{eqcenh}). The latter is indeed an excellent approximation of the excited state probability $c_e(t)$ in a coupling to a sufficiently large FQC (see Fig.\ref{figure6}). We find

\begin{equation}
c_{k}(\infty)=\frac{v \Omega_0}{\hbar \sqrt{\Delta_0}}\bigg[\frac{-1}{-\frac{\Gamma}{4}+\frac{i\sqrt{\Delta_0}}{2}+ik \delta}+\frac{1}{-\frac{\Gamma}{4}-\frac{i\sqrt{\Delta_0}}{2}+ik \delta}\bigg]
\end{equation}
with $\Delta_0=4\Omega_0-\frac{\Gamma^2}{4}$. 
Figure~\ref{figure8} shows the probability occupations $|c_k(t_f)|^2\simeq |c_{k}(\infty)|^2$. These distributions  exhibit two sidebands centered about $k$ values such that $ |k| \delta \simeq \Omega_0$, symmetrically distributed around $k=0$ for our choice $\Delta=0$. A similar generation of sidebands is observed in the dynamical Casimir effect~\cite{DynamicalCasimirReview}. These occupancy probabilities actually determine the number of relevant FQC states and the size of the minimal appropriate FQC. 
Indeed, we have indicated on Figs.~\ref{figureII3}(a,b,c) the maximum occupancies number $N_{\rm max}(v)$ as a function of the coupling strength $v$. This quantity is defined as $|c_{N_{\rm max}(v)}(t_f)|^2 = {\rm \max}_n \{ |c_{n}(t_f)|^2  \}$ for the considered coupling strength $v$ and Rabi frequency $\Omega_0$. As the occupation peak corresponds approximately to the Rabi frequency, we expect  $N_{\rm max}(v) \simeq \Omega_0 /\delta = \Omega_0 \hbar^2 \Gamma / (2 \pi v^2)$ from Eq.~\eqref{eq1}. In Figs.~\ref{figureII3}(a,b,c), the line representing the maximum occupancy number $N_{\rm max}(v)$ is almost superimposed with the interface between the suitable and unsuitable FQCs (white/grey zones respectively). This confirms that the suitable FQC are those that host all the significantly populated levels. The population of each Fourier components is represented for different $\Omega_0/\Gamma$ ratios in Fig.~\ref{figure8}: in the weak coupling limit $\Omega_0 \ll \Gamma$, $N_{\rm max}(v)$ is mainly determined by the dissipation rate $\Gamma$, while in the strong coupling limit  $\Omega_0 \gg \Gamma$, it scales linearly with the Rabi frequency $\Omega_0$ (for $v=0.3$, $N_{\rm max}(v) \simeq 1.77 \Omega_0$).

\section{Non-Markovian dynamics and adaptive quasi-continuum}
\label{Sec4}

In this Section, we analyse quantitatively finiteness-related effects in the evolution of a FQC-coupled quantum system.  First, we  establish a connection between the presence of revivals (discussed in Section~\ref{Sec2}) and a measure of non-Markovianity applied to the FQC-emulated dynamics. Secondly, we show that the FQC structure of equidistant energy levels induces a mismatch of the effective Rabi frequency and decay rates when compared to the equivalent parameters in the non-Hermitian model. Solving this issue suggests an adaptative structure of FQCs, discussed below, capable of reproducing non-Hermitian dynamics with a considerably reduced number of states.

\subsection{Revivals and non-Markovianity}

Revivals in the excited state probability $\pi_e(t)$, discussed in Sec.~\ref{Sec2} for the single level system, also occur in the two-level FQC emulated dynamics for large values of the coupling strength $v$. Such revivals are indeed a symptom of non-Markovian dynamics in the FQC: their exact form depends on the initial quantum state and therefore reveals a memory effect in the quantum evolution.  Despite the successful emulation of dissipative dynamics over a given time interval, these  revivals show that some information on the initial state has been transmitted and stored in the FQC. The revival appears as a kind of constructive interference effect when information about the initial state, stored in the FQC, returns to the system. Non-Hermitian dynamics~\eqref{eq:nonHermitianexact} are Markovian, and so the emergence of non-Markovianity reveals a discrepancy between the FQC-emulated system and the ideal irreversible case.

These considerations suggest studying quantitatively the non-Markovianity of the FQC-emulated dynamics. We proceed by using the measure of Ref.~\cite{Breuer09}, summarized below for convenience. This measure uses the trace distance $T(\rho,\sigma)$~\eqref{eq4}, which has a direct interpretation in terms of the distinguishability of the associated quantum states. Indeed, if we consider an emitter which randomly prepares one of the two quantum states $\{ \rho,\sigma \}$ with equal probability, the probability of an observer to successfully identify the correct quantum state by a measurement is simply $\frac 1 2(1 +T(\rho,\sigma))$. Markovian processes correspond to a decreasing trace distance for any set of states following the quantum evolution associated with the process. In this case, no information likely to improve the dinstinguability of the states $\{ \rho(t),\sigma(t) \}$ is acquired by the system during the evolution. The unitary evolution operator of a closed quantum system, and more generally complete positive trace-preserving maps, fall into this category.  Conversely, non-Markovian quantum processes are those that exhibit at least a temporary positive variation of the trace distance for some pair of initial states. This increase witnesses a flow of information from the environment back to the system.

\begin{figure}[h!]
    \centerline{\includegraphics[scale=0.68]{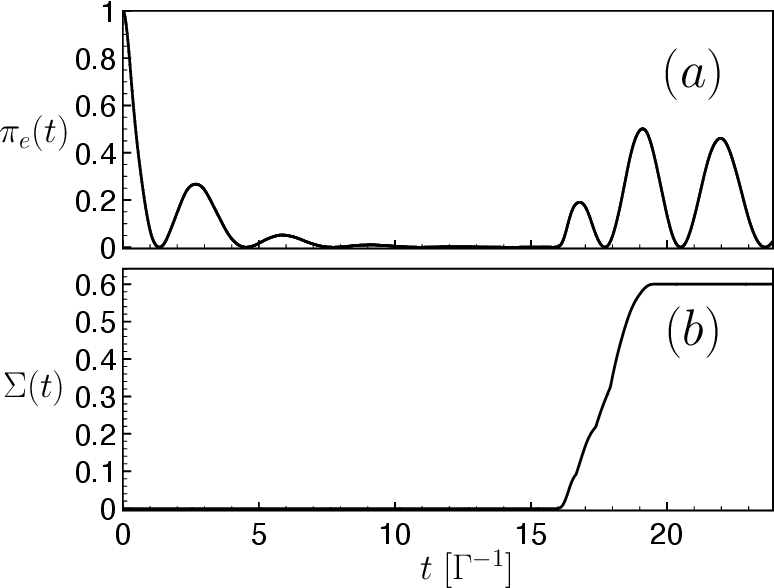}}
    \caption{Revivals and non-Markovianity: (a) Time evolution of the population in the unstable state $\pi_e(t)$. (b)Measure of the non-markovianity $\Sigma(t)$ (see text) of a two level system composed of $2N+3=53$ levels as a function of time. Parameters: $\Omega_0=1 \: \Gamma$, $v=0.25 \: \hbar \Gamma$. The non-Markovianity measure is estimated within a $1\%$ accuracy by a sampling over a set of $256$ initial states.}
    \label{figure9}
\end{figure}

 To obtain a quantitative measure, one introduces the rate of variation of the trace distance for a given quantum process
\begin{equation}
	\sigma_{\rho_{1}^0,\rho_{2}^0}(t) = \frac{d}{dt} T(\rho_1(t),\rho_2(t)).
\end{equation}
where $\rho_{1,2}(t)$ are two density matrices undergoing the quantum process under consideration and therefore following the same evolution operator/dynamic equation, but with distinct initial conditions $\rho_{1,2 \:}(0)= \rho_{1,2}^0$. Quantum processes with $\sigma_{\rho_{1}^0,\rho_{2}^0}(t) >0$ correspond to an increasing trace distance, and therefore a flow of information from the environment to the system.
The non-Markovianity measure is given by~\cite{Breuer09}
\begin{equation}
	\Sigma(t) = {\rm max}_{\rho_{1}^0,\rho_{2}^0} \int_0^{t} dt'  \: \Theta( \sigma_{\rho_{1}^0,\rho_{2}^0}(t') ) \: \sigma_{\rho_{1}^0,\rho_{2}^0}(t')
\end{equation}
The Heaviside function $\Theta(x)$ (s.t. $\Theta(x)=1$ if $x \geq 0$ and $\Theta(x)=0$ for $x < 0$) guarantees that only time intervals with an increasing trace distance effectively contribute to the integral. The quantity $\Sigma(t)$ is obtained by considering all possible initial quantum states $\rho_i^0=| \psi_i \rangle \langle \psi_i| $ (with $|\psi_i \rangle$ a generic quantum state of the full Hilbert space), and the considered evolution corresponds to $\rho(t)= P_{\rm eg} e^{- i H t /\hbar} \rho^0 e^{ i H t /\hbar} P_{\rm eg}$, where $H$ is the Hamilonian~\eqref{Hal2} and $ P_{\rm eg}$ the projection operator introduced earlier on the two-dimensional subspace. In Fig.~\ref{figure9}b, we have plotted the evolution of the non-Markovianity $\Sigma(t)$ as a function of time for a given FQC, to be compared with the time evolution of the excited-state population $\pi_e(t)$ on Fig.\ref{figure9}a. We have deliberately chosen a time window during which a revival is observed. Figures~\ref{figure9}(a,b) reveal that a sharp increase in the non-Markovianity $\Sigma(t)$ does occur at the onset of the probability revival. The non-Makovianity $\Sigma(t)$ thus provides another determination of the time window over which the FQC dynamics accurately emulates an irreversible non-Hermitian evolution.

\subsection{Adaptive FQC}
\label{Sec52}

The study carried out in Sec.~\ref{Sec3} reveals that the minimal size of suitable FQCs scales with the Rabi frequency $\Omega_0$. Here we go a step further, and propose to adapt the FQC's structure depending on the Rabi coupling $\Omega_0$. We no longer consider exclusively flat FQCs with equidistant levels around the excited state energy. Instead, we study adaptive FQCs with an enhanced density of states  around the occupation peaks depicted in Fig.~\ref{figure8}. As seen below, such adaptive FQCs yield an optimized emulation of non-Hermitian dynamics.

\begin{figure}[t!]
    \centerline{\includegraphics[scale=0.65]{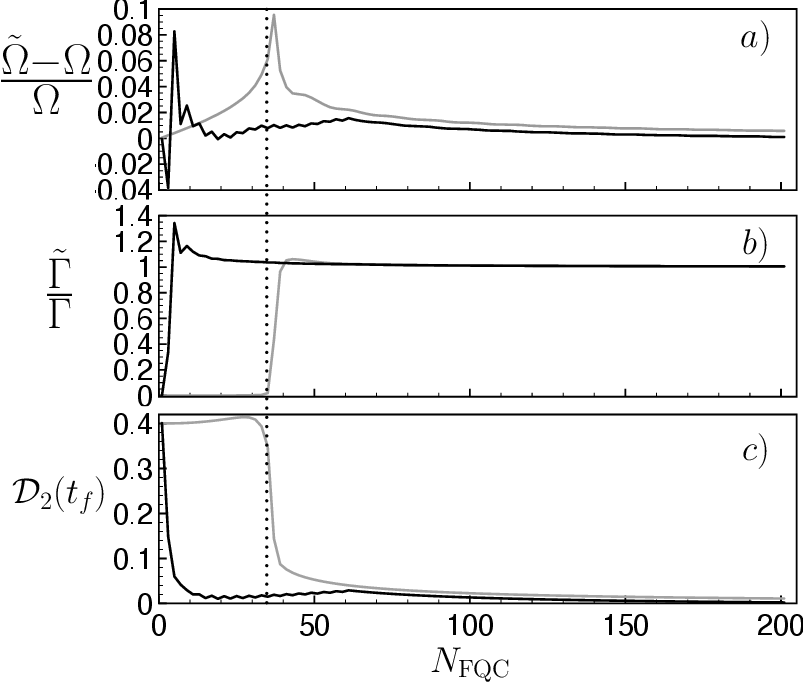}}
    \caption{Mismatch between the effective Rabi frequency (a), damping rates (b), and mean trace distance (c) as a function of the size $N_{\rm FQC}$ obtained by comparing the FQC model with the non-Hamiltonian model. We have plotted the effective parameters obtained from a flat FQC made of equidistant levels (gray solid line) and for an FQC with an adaptative structure with removed central states (black solid line). The dotted line corresponds to $N_{\rm FQC}=35$, considered in Fig.\ref{figure11}. Parameter $v=0.3 \: \hbar \Gamma$.}
    	    \label{figure10}
\end{figure}

We begin by investigating the influence of the discrete FQC structure on the emulated quantum dynamics. Figure~\ref{figureII3}c exhibits a slightly gray zone associated with a slight mismatch between the FQC evolution and the non-Hermitian dynamics. This is the regime we wish to investigate. For this purpose, we consider non-Hermitian dynamics~(\ref{eqcenh})
in the strong coupling regime $\Omega_0 \gg \Gamma$. The corresponding excited-state population reads
\begin{equation}
	\pi_e(t) = e^{-\Gamma t/2}\cos^2 (\Omega t),
	\label{equaform}
\end{equation}
with $\Omega=\Omega_0 (1-(\Gamma/4 \Omega_0)^2)^{1/2}$. An effective Rabi frequency $(\tilde{\Omega})$ and dissipation rate $(\tilde{\Gamma})$ for the FQC dynamics are obtained by fitting the excited-state population $\tilde{\pi}_e(t)$ with the form~\eqref{equaform} of exact non-Hermitian dynamics. The corresponding results are represented as a solid gray line in Fig.~\ref{figure10}(a,b) for different FQCs.

The discrepancy between the FQC model and the ideal non-Hermitian case can be explained by a closer examination of the integration Kernel~\eqref{eqce}, or more precisely its equivalent for the two-level case. To reproduce non-Hermitian dynamics, the integration Kernel must take a form analogous to Eq.~\eqref{eqce2}.  In this case, the frequency mismatch $\Delta \Omega=\tilde \Omega-\Omega$ cannot be attributed to a Lamb shift effect, as the principal part of the Kernel in Eq.~\eqref{eq6} cancels out in the presence of a symmetric FQC with homogeneous coupling constant. The slight frequency shift is therefore a signature of the non-Markovianity of the FQC dynamics, i.e. of the residual error committed by replacing Eq.~\eqref{eqce} by Eq.~\eqref{eqce2}. The corresponding approximations, namely of short Kernel memory and the extension of the integral in Eq.~\eqref{eqce2} to infinity,
are indeed jeopardized by the discrete FQC structure. Intuitively, the discrete states of zero-energy (i.e. of energy close to the unstable excited state) can increase the error. We  also note that these central states are not significantly populated in the FQC dynamics: the highly populated levels correspond to peaked population sidebands centered on $\pm \Omega_0/\delta$.

\begin{figure}[t!]
    \centerline{\includegraphics[scale=0.70]{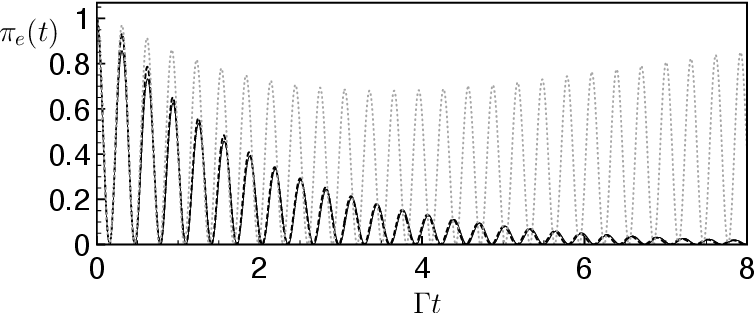}}
    \caption{FQC-emulated dynamics in the strong coupling regime ($\Omega_0=10 \Gamma$): excited population as a function of time for a flat FQC (dotted gray line) made of $N_{\rm FQC}=2N+1=35$ equidistant levels and for an adaptative FQC (dashed black line) made of $N_{\rm FQC}=2N=34$ levels. For both FQCs we have used the parameters $v=0.3 \: \hbar \Gamma$. The solid gray line represent the exact evolution expected from the non-Hermitian dynamics (Eq.\eqref{equaform}), and is almost superimposed to the adaptative FQC results.
    }
    \label{figure11}
\end{figure} 
These observations raise the question of the relevant optimal FQC structure in this regime. From Fig.~\ref{figureI3}(b), the central FQC state eigenenergies undergo the largest shift from the linear dispersion relation expected from an ideal continuum. Furthermore, Fig.~\ref{figure8}(c) shows that in the strong driving regime ($\Omega_0 \gg \Gamma)$, the final population of these states is very small. These considerations suggest that the central components of the FQC play a minor role, or even a deleterious role.

 To confirm this intuition, we studied a different FQC model obtained from the former FQC by removing the states close to the $E=0$ energy while preserving the symmetry of the distribution. The corresponding results are shown in Figs.~\ref{figure10}(a,b,c) (solid black line). For large FQC sizes, both the flat and adaptive FQC provide a good emulation of non-Hermitian dynamics,  although the latter achieves the same error in the $\tilde \Omega, \tilde \Gamma, \mathcal{D}_2 $ parameters with a much smaller size. For small sizes $N_{\rm FQC} \leq 30,$ the spectrum of the regular flat FQC is too narrow to include the highly populated Rabi sidebands of Fig.\ref{figure8}(c). Consequently, small regular FQCs produce a negligible effective damping rate $\tilde \Gamma$. On the other hand, the adaptive FQC contains by construction states nearby these sidebands. Thus, even small adaptive FQCs $N_{\rm FQC} \simeq 10$ already give an effective damping rate $\tilde \Gamma$ close to the appropriate value. At intermediate sizes ($10 \leq N_{\rm FQC} \leq 50$) adaptive FQCs also outperform regular FQCs: a strong improvement is observed in the agreement between the effective  Rabi frequency $\tilde \Omega$ and damping rate $\tilde \Gamma$ with their non-Hermitian counterparts $\Omega,\Gamma$ as well as a significant reduction of the mean trace distance compared to exact non-Hermitian dynamics. We conclude that, for the damped Rabi dynamics considered, adaptive FQCs with a tailored distribution (involving mostly states close to the Rabi frequency sidebands $\pm \Omega_0/\delta$ and presenting a hole in the central zone nearby the unstable state energy ($E=0$)) provide a higher fidelity to non-Hermitian dynamics with constant resources, i.e. with the same number of states and for an identical time-window .

 Figure~\ref{figure11} provides an example where both kind of FQCs (flat vs adaptive) produce very different qualitative behaviors while having a very similar number of states. While the coupling to a regular equidistant FQC cannot account for the damping of the Rabi oscillation, the quantum system coupled to the adaptive FQCs yields an excellent agreement with the predicted non-Hermitian dynamics~(Eq.\eqref{equaform}).

\section{Conclusion}

In conclusion, we discussed the emulation of non-Hermitian quantum dynamics during a given time window with a finite quasi-continuum composed of discrete states. We specifically considered  the exponential decay of an unstable state, and the Rabi driving of a two-level quantum system exhibiting an unstable state. We have characterized the short and long time deviations of the FQC-emulated system compared to the exact non-Hermitian case. Short-time deviations can be interpreted in terms of the Zeno effect, while the long-term deviations correspond to a probability revival that can be quantified by a measure of non-Markovianity. We have provided a criterion for the adequacy of the  discrete FQCs considered by evaluating the occupancy probabilities of the quasi-continuum states. There is a trade-off between using FQCs involving a large number of states and achieving high accuracy in emulating non-Hermitian dynamics. We have shown that in the strong coupling regime, this trade-off can be significantly improved by considering FQCs with an adapted density of states.   This study is potentially relevant for many-body systems, where a given subsystem can be coupled to a large set of states corresponding to the surrounding bodies~\cite{QuantumSimulatorRoadmap}. Quantum dots coupled to nano-wires are a promising platform to implement low-dimensional systems coupled to FQCs~\cite{Ricco22A,Ricco22B}. This work also paves the way for the emulation of non-Hermitian dynamics with a finite set of states. A long-term goal is to integrate a tunable dissipation within quantum simulators~\cite{QuantumSimulatorRoadmap}. Different methods have been investigated to reach this goal, relying on the Zeno effect \cite{Ref9,Ref10,Ref11,Ref12}, atom losses \cite{Ref15,Ref16}, or multichromatic Floquet \cite{FIDGO23} to name a few.

\bibliography{biblio2_FI}

\end{document}